# Thinking Outside the Band: Absorptive Filtering

Matthew A. Morgan

**Introduction**

Today's high-frequency radio system engineer has at his fingertips an encyclopedic body of work to draw upon for his filtering requirements – from topologies, to synthesis equations, to proven methods of implementation. Compiled from decades of work by countless researchers, the canon of literature on filters nevertheless focuses almost exclusively on techniques that assume from the outset that all circuit elements will be nominally-lossless (some few exceptions are noted in [1]-[9]). This constraint serves most of the conventional ideas about what a filter should be quite well, including the minimization of passband insertion loss and the sharpest possible transition between passband and stopband. Fundamentally, however, it demands that the out-of-band power rejected by the filter is reflected back to its source. In microwave terminology, the impedance mismatch is maximized. In a field where impedance-matching is almost as automatic as breathing, this departure from the norm is accepted because it only occurs out of the operating band.

It is easy to fall into a pattern of thinking that in-band performance is all that matters, and that engineers should not care how their circuits behave outside of the nominal operating frequency range. This is not always true. (If out-of-band signal power did not matter, why filter it in the first place?) Frequency conversion circuits, like mixers and multipliers, may respond in unpredictable ways to chance reactive terminations at their ports outside of the nominal input and output frequency ranges – hence the common practice of "padding" the RF and IF ports of mixers, not only to improve an often mediocre in-band impedance match, but also to desensitize them to broadband impedance variations. Out-of-band noise and spurious CW tones may be up- or down-converted into the operating band, degrading signal-to-noise ratio and dynamic range. Further, amplifier gain rarely drops off as sharply as one would like outside of the passband, especially now with the proliferation of wide-band and multi-band consumer electronics, giving rise to excess power at uncontrolled frequencies. Conventional reflective filters do nothing to dissipate these spurious signals or added noise. Instead they build up as standing waves between the filter and the source and/or are radiated into the electronics housing, impacting the linearity, stability, and electromagnetic compatibility of the system. These effects often go unnoticed, but they can be important in some high-performance applications.

For these reasons, it is sometimes advantageous to design a filter that absorbs the stopband spectrum rather than reflecting it. As this requires at least some lossy elements to be used, care must be taken to ensure that the low in-band insertion loss and impedance-matching are not discarded in the process. The two most common ways of doing this are with balanced filters – two identical, conventional filters coupled at the input and output with 3 dB quadrature hybrids – and diplexers where the output of one or more branches is terminated with a resistor. There are, however, far simpler frequency-selective circuits that possess the desired broadband impedance match, while also exhibiting greater amplitude- and phase-stability with extended frequency applicability.

**Reflectionless Filter Cell**

The generic reflectionless filter cell with which we will be concerned is shown in Fig. 1. It is a symmetric network derived by the application of even- and odd-mode analysis with the constraint that the resultant two-port will have identically zero input reflection coefficient at all frequencies [10]-[11]. The structure shown is low-pass, but may be trivially converted to any other configuration – high-pass, band-pass, band-stop, and multi-band – by well-known transformation techniques [10]. Since all of the salient features of this topology translate directly into those other domains, the discussion in this article will be confined to low-pass filters for simplicity.

To illustrate, let us consider the transfer characteristic of the single-pole cell shown at the top of Fig. 2. We may begin by decomposing the excitation of a single port into two simultaneous excitations of both ports, one in which both ports are excited in phase with one another, and the other with both ports excited in anti-phase. By superposition, the net response of the network to the single-sided excitation must be a linear combination of these two. Further, when the two ports are excited in phase, a virtual open-circuit may be drawn down the symmetry plane, and the entire circuit response may be determined by a reduced circuit containing at most half of the elements. This is called the *even-mode equivalent circuit*. Conversely, when the two ports are excited in anti-phase, a virtual short-circuit may be drawn down the symmetry plane to arrive at the *odd-mode equivalent circuit*. The input impedance of these two equivalent circuits is given by

$$Z_{even} = \tfrac{1}{j\omega C} + \left(\tfrac{1}{j\omega L} + \left(R + \tfrac{1}{j\omega C}\right)^{-1}\right)^{-1} \tag{1a}$$

$$Z_{odd} = \left(\tfrac{1}{j\omega L} + \left(\tfrac{1}{j\omega C} + \left(\tfrac{1}{j\omega L} + \tfrac{1}{R}\right)^{-1}\right)^{-1}\right)^{-1}. \tag{1b}$$

To simplify the algebra, let us substitute $L = Z_0/\omega_p$, $C = Y_0/\omega_p$, and $R = Z_0$, and normalize the frequency variable to the pole frequency, $f = \omega/\omega_p$. The above expressions then become

$$Z_{even} = Z_0\left(\tfrac{1}{jf} + \left(\tfrac{1}{jf} + \left(1 + \tfrac{1}{jf}\right)^{-1}\right)^{-1}\right) = Z_0 \tfrac{1 - 2f^2 + jf(1-f^2)}{-f^2 + jf(1-f^2)} \tag{2a}$$

$$Z_{odd} = Z_0\left(\tfrac{1}{jf} + \left(\tfrac{1}{jf} + \left(\tfrac{1}{jf} + 1\right)^{-1}\right)^{-1}\right)^{-1} = Z_0 \tfrac{-f^2 + jf(1-f^2)}{1 - 2f^2 + jf(1-f^2)}. \tag{2b}$$

The reflectionless coefficients of the even- and odd-mode equivalent circuits are then given by

$$\Gamma_{even} = \tfrac{Z_{even} - Z_0}{Z_{even} + Z_0} = \tfrac{1-f^2}{1 - 3f^2 + j2f(1-f^2)} \tag{3a}$$

$$\Gamma_{odd} = \tfrac{Z_{odd} - Z_0}{Z_{odd} + Z_0} = \tfrac{-(1-f^2)}{1 - 3f^2 + j2f(1-f^2)}. \tag{3b}$$

The full two-port scattering parameters may then be recovered as follows

$$s_{11} = s_{22} = \tfrac{1}{2}(\Gamma_{even} + \Gamma_{odd}) = 0 \tag{4a}$$

$$s_{21} = s_{12} = \tfrac{1}{2}(\Gamma_{even} - \Gamma_{odd}) = \tfrac{1-f^2}{1 - 3f^2 + j2f(1-f^2)}. \tag{4b}$$

The reflectionless property is thus proven by (4a). The transfer characteristic, (4b), is plotted in Fig. 3. The perfection of the impedance match is of course theoretical and dependant on the quality and tolerance of the elements used. Measured circuit performance for low-pass and band-pass prototypes consisting of multiple cascaded cells is shown in [10]. The stop-band peak occurs at $f=\sqrt{3}$ (relative to the pole frequency) and has a value of $(4 + j2\sqrt{3})^{-1}$, limiting stop-band rejection per cell to $10\cdot\log(28) = 14.47$ dB.

The poles and zeros may be identified simply by substituting the complex frequency $f = -js$ and factoring the transfer function (4b),

$$H(s) = \tfrac{(s-j)(s+j)}{(s+1)\left(s+\tfrac{1+j\sqrt{7}}{4}\right)\left(s+\tfrac{1-j\sqrt{7}}{4}\right)}. \tag{5}$$

Thus, the poles are at $s = -1$ and $(-1\pm j\sqrt{7})/4$, and the zeros are at $s = \pm j$. It can be shown, with the proper frequency scaling, that this pole-zero configuration corresponds to that of a third-order Chebyshev Type II (or Inverse Chebyshev) filter with ripple factor $\varepsilon = (3\sqrt{3})^{-1} \approx 0.192$. [12]

An individual reflectionless filter cell may not have sufficient stopband rejection on its own to satisfy system requirements, but the cumulative effect of several in cascade can be competitive with conventional filter types, especially in scenarios where broad impedance-matching is required. Consider, for example, the transfer characteristic of four single-pole cells when compared to matched (resistively-terminated) diplexers based on $10^{th}$-order Chebyshev (Type I) and $16^{th}$-order Butterworth prototypes, as shown in Fig. 4. A quality factor of $Q = 30$ was assumed for the inductors. Note that the overall frequency selectivity (defined in this example as the difference in 3 dB and 60 dB corners, respectively) is the same in all cases.

The performance of higher-order cells may be derived in a similar manner, and in general result in sharper cutoff at the expense of poorer stopband rejection at the peak [10]. Thus, arbitrarily sharp cutoff is available, but larger numbers of cells must be cascaded to achieve a given level of attenuation in the stopband.

**Distributed Filtering**

Though it usually occurs without pre-meditation, the standard practice for RF designers is to build-up a system using relatively broadband components and then define the operating band with a single filter, in effect attenuating all of the accumulated out-of-band noise, interferers, and spurious tones all at once. In that sense, it would be preferable to place the filter at the very end of the analog portion of the system, or just prior to a frequency conversion, so that all unwanted signals and sideband noise may be filtered out before they are digitized or irrevocably overlaid upon an image band. On the other hand, dynamic range and linearity requirements tend to encourage filtering earlier in the RF or IF chain so that sensitive front-end components are not overdriven. To get the best of both worlds, one would like to have filters in several places, but this is usually an inefficient and costly practice considering the relative complexity of conventional high-order filters, as well as the potential for negative out-of-band interactions between the filter and neighboring components discussed previously.

In this light, the small, divisible cells of the reflectionless filters represent a substantial advantage. Since each of the cells are small and individually matched, they need not be adjacent in the signal path. They may, instead, be distributed throughout the system as needed without risk of poor out-of-band interactions. On the contrary, when used in this manner they usually ameliorate other subtle out-of-band effects that are not always well-understood, resulting in a less glitchy, more stable system. They also prevent out-of-band signals from accumulating or getting amplified, keeping the unwanted signals at a lower level throughout the chain rather than allowing them to build and then beating them back down in the final step.

**Stability**

Apart from the system-stabilizing effects of good broadband impedance-matching described above, the reflectionless filter itself is also more stable in amplitude and phase than its conventional counterparts, especially near the edge of the band, which can be a boon to precision applications where system calibration and component drift is critical. This superior stability is evident from the derivatives of the complex gain (4b) when contrasted with that of traditional filter topologies [10], but is perhaps illustrated more clearly by an example.

Consider the reflectionless and conventional Chebyshev filters in Fig. 4. A yield analysis was performed to elucidate the gain changes that result from small variations in the component values. The result is shown in Fig. 5. This plot shows the magnitude of the net change in filter transfer characteristic (amplitude and phase) between 25 C and 35 C over a normal distribution of component values, each having 2% tolerance and a temperature coefficient of 50 ppm/°C. Not only does the complex gain of the reflectionless filter change less with temperature (it has a lower peak), it is also much more consistent (less spread). As expected, the bulk of the variation for both filters is concentrated near the band edge.

**Frequency Versatility**

For any given manufacturing process, there is a limited range of component values available, and these components are limited in the frequency range over which they can be used effectively. Surface mount components, for example, can only be made so small in value before the pad capacitance and/or trace inductance dominates their behavior, and can only be made so large in value before losses, self-resonant effects, and/or dielectric breakdown render them useless. MIM capacitors that are too small suffer from excessive tolerance error due to fringing and contributions from the coupling traces, whereas those that are too large become overmoded. Spiral inductors, transmission lines, thin-film resistors – components of every variety – all are subject to upper and lower bounds introduced by the manufacturing process.

A curious feature of the reflectionless filter cell as drawn in Fig. 1 is that all elements of a single type – resistors, capacitors, and inductors – have the same value, no matter what the order. This means the reflectionless filter will generally require more moderate values, neither too big nor too small, for a given passband than that of its conventional counterparts. To highlight the point, consider that the Chebyshev diplexer in Fig. 4 required capacitors ranging from 6.5-31 pF, and inductors from 21-103 nH. The Butterworth diplexer required 1.6-55 pF, and 12-414 nH, respectively. The reflectionless filter, however, required only 9.2 pF capacitors, and 23 nH inductors.

This reduced component spread will enhance frequency versatility in at least two ways. First, the larger components required by conventional filter topologies when compared to the reflectionless filter cell will tend to have larger parasitics and lower self-resonant frequencies. On the other end of the scale, the smallest components required as you move to higher frequencies may simply be unavailable, and in any case become lost in the parasitics of the manufacturing process.

One might argue that the moderate component values are a consequence of building up the composite filter by cascading several low-order filters rather than a single high-order filter. There is truth in this argument, however the comparison is not entirely fair, because the reflectionless property is what allows them to be cascaded so effectively;

conventional filters cannot be cascaded in this way without severe passband distortion due to mismatches in the transition region. (And again, even the higher-order reflectionless filter cells in Fig. 1, which do have sharper cutoffs, still require only the single component values of each type.)

**Conclusions**

The reflectionless filter cell described in this article alleviates many system problems associated with excess out-of-band gain, impedance mismatches, and component interactions. The simplest filter cell exhibits a third-order Inverse Chebyshev response with 14.47 dB peak stopband attenuation, and can be cascaded for additional attenuation as needed. They are simple to design and easy to use – for much like small fixed attenuators ("pads"), they can be placed anywhere in the signal path desired without fear of causing unwanted standing waves, and instead reducing them where they already exist beyond the intended frequency range. Compared to conventional filter topologies with similar cutoff frequencies, they exhibit an order of magnitude greater amplitude and phase stability, ensuring accurate and repeatable complex gain in calibrated systems. Finally, the component requirements are moderate in value and minimal in number, easing their implementation while improving design yield, and extending the applicability of a given component technology beyond the normal range, both above and below.

**Figures**

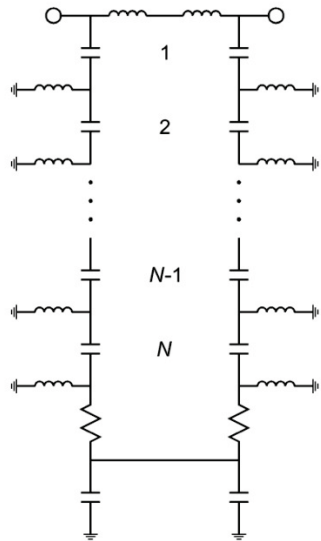

Fig. 1.  Low-pass reflectionless filter cell. As drawn, all resistors, capacitors, and inductors, respectively, have the same value for a given cutoff frequency.

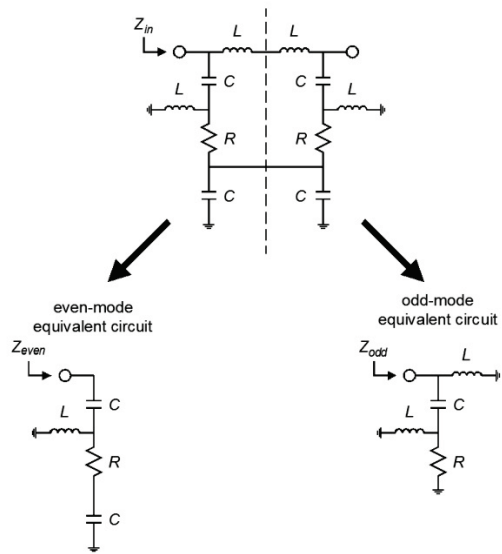

Fig. 2.  Analysis of a single-pole reflectionless filter cell by even- and odd-mode equivalent circuits.

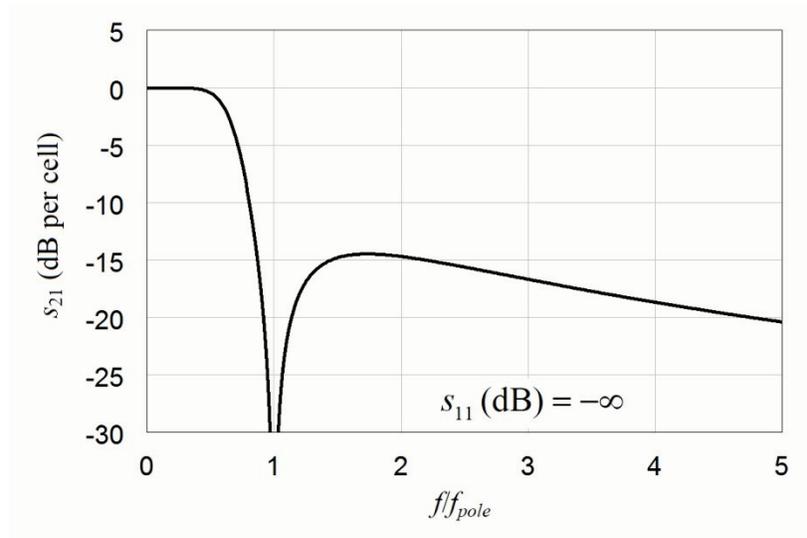

Fig. 3.   Theoretical response of a single-pole low-pass reflectionless filter cell.

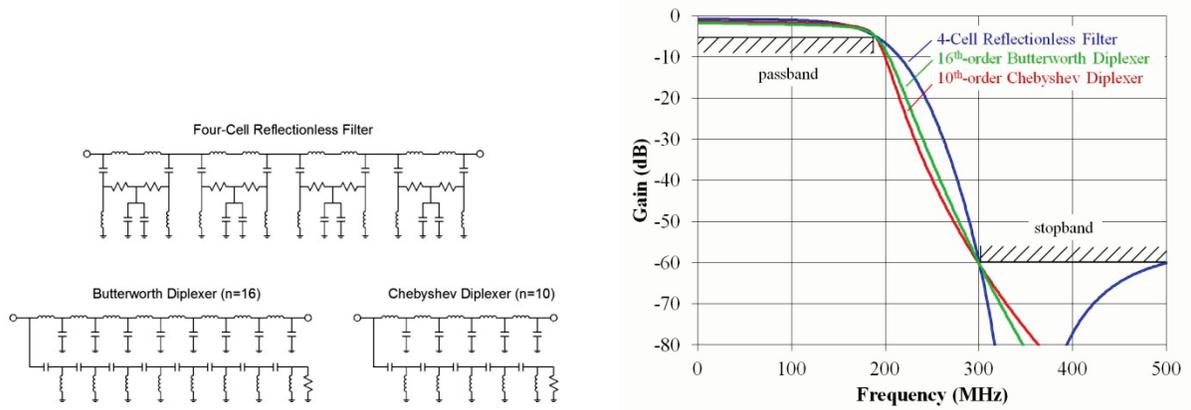

Fig. 4.   Frequency selectivity of a cascade of 4 reflectionless filter cells, compared to 10[th]-order Chebyshev and a 16[th]-order Butterworth diplexers. Inductor $Q = 30$.

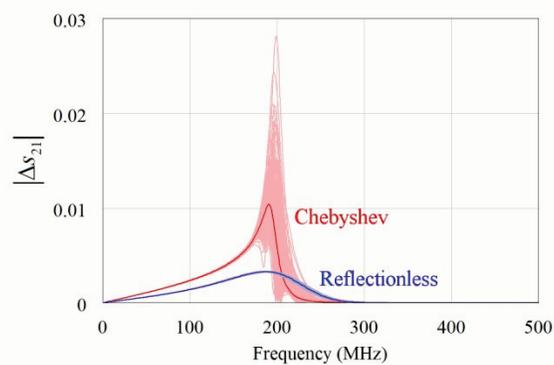

Fig. 5.   Yield simulation showing the stability of the transmission coefficient for 10[th]-order Chebyshev and 4-cell Reflectionless filters.